\documentclass{svjour2}                    
\smartqed  
\usepackage{amssymb}
\usepackage{graphicx}
\usepackage{slashed}
\usepackage{psfrag}
\usepackage{rotating}
\begin{document}

\title{Study of solid $^4$He in two dimensions}
\subtitle{The issue of zero--point defects and study of confined crystal}

\titlerunning{Study of solid $^4$He in 2D}        

\author{M. Rossi \and L. Reatto \and D.E. Galli}

\authorrunning{M. Rossi \and L. Reatto \and D.E. Galli} 

\institute{M. Rossi \at
Dipartimento di Matematica, Politecnico di Milano, Piazza Leonardo da Vinci 32, 20133 Milano, Italy\\
           \and
           L. Reatto \at
Dipartimento di Fisica, Universit\`a degli Studi di Milano, via Celoria 16, 20133 Milano, Italy\\
           \and
           D.E. Galli \at
Dipartimento di Fisica, Universit\`a degli Studi di Milano, via Celoria 16, 20133 Milano, Italy
}

\date{Received: date / Accepted: date}

\maketitle

\begin{abstract}
Defects are believed to play a fundamental role in the supersolid state of $^4$He.
We report on studies by exact Quantum Monte Carlo (QMC) simulations at zero temperature of the
properties of solid $^4$He in presence of many vacancies, up to 30 in two dimensions (2D).
In all studied cases the crystalline order is stable at least as long as the concentration of
vacancies is below 2.5\%.
In the 2D system for a small number, $n_v$, of vacancies such defects 
can be identified in the crystalline lattice and
are strongly correlated with an attractive interaction.
On the contrary when $n_v \gtrsim 10$ vacancies in the relaxed system disappear and in their place
one finds dislocations and a revival of the Bose-Einstein condensation.
Thus, {\it should} zero--point motion defects be present in solid $^4$He, such defects would be
dislocations and not vacancies, at least in 2D.
In order to avoid using periodic boundary conditions we have studied the exact ground state
of solid $^4$He confined in a circular region by an external potential.
We find that defects tend to be localized in an interfacial region of width of about 15 \AA.
Our computation allows to put as upper bound limit to zero--point defects the concentration
$3\times10^{-3}$ in the 2D system close to melting density.
\keywords{zero--point defects \and solid $^4$He \and supersolid}
 \PACS{67.80.-s \and 67.80.bd \and 67.80.dj}
\end{abstract}

\section{Introduction}
\label{intro}

Supersolidity is an intriguing state of matter in which spatial order, typical
of the solid phase, and off-diagonal long range order \cite{Penr} (ODLRO), which characterizes
the Bose--Einstein condensation (BEC) phenomena, are simultaneously present, implying
some form of superfluid properties even for the solid phase.
This striking idea of a superfluid solid was proposed long ago \cite{Andr,Ches} and 
solid $^4$He was early recognized as the natural
candidate to display such a counterintuitive coexistence of orders.
This topic
attracted the attention of many physicists \cite{Meis} before going out of attention in the
absence of experimental evidence.
As a consequence of the recent discovery of non classical rotational inertia 
(NCRI) \cite{Chan} in solid $^4$He,
one of the expected manifestations of supersolidity \cite{Legg}, the 
possible existence of a supersolid phase has recently gained once more the attention of the
scientific community, making solid $^4$He systems the subject of many experimental and 
theoretical studies (See Refs. \cite{Prok,Bali,Revi,Bal2} for recent reviews).

There is strong evidence that defects play an important role in NCRI.
In fact, experiments show that usually NCRI is strengthened by increasing disorder in the crystal, and 
microscopic simulation studies agree on the fact that an ideal perfect crystal
does not show ODLRO \cite{n0T1,n0T2} even at $T=0$ K \cite{Vita,Revi}.
By {\it ideal perfect crystal} we mean a crystalline solid extended to all space with {\it one atom per lattice site}.
A key question is if bulk solid $^4$He is an ideal perfect crystal or not.
For instance, if zero--point independent vacancies were present, the crystal would still be perfect
because vacancies are delocalized defects, but not ideal because there would be less than one atom
per lattice site (this is also called an incommensurate solid).
Notice that the earlier proposal \cite{Andr,Ches} for the supersolid state assumed the presence of
zero--point defects in the ground state of solid $^4$He in the form of ground state vacancies.
Presently there is a prevalent view that supersolidity is an extrinsic
property of solid $^4$He. However not all share this view \cite{Ande,Revi,Ande2}
and present experiments cannot rule out the presence of intrinsic defects\cite{Sim1,Sim2}.
Moreover, it is still not clear which kind of disorder contributes to,
or is responsible for, the anomalous properties observed in solid $^4$He.
Many different defects have been considered, ranging from vacancies to grain boundaries,
passing from dislocations to quantum glasses; but none of the proposed models seems
able to capture the whole phenomenology of supersolidity in solid $^4$He \cite{Bal2}.
Thus, independent of the very relevant issue of intrinsic or extrinsic defects,
the knowledge of the properties of defects in this system is one of the
main goals of the present theoretical investigations on supersolidity in this system.

From the theoretical point of view the method of choice to investigate
a strongly interacting quantum system like solid $^4$He is Quantum Monte Carlo (QMC) simulations.
Nowadays a number of studies on defected solid $^4$He systems studied by means 
of QMC techniques are present in literature
\cite{Pede,Gallo,Gall1,spig1,spig2,spig3,Gall,Boni,Cepe2,Pollet,Pollet2,Boni2,Corb,Soyl,Vit1,Vit2,Fate,Bor1,Bor2,Bor3}.
Crystals with defects can be simulated with an appropriate choice of the simulation box (SB) and the
number of particles such that, combined with the use of periodic boundary conditions (PBC),
the MC sampling is constrained to configurations which host the desired defect.
When PBC are not able to stabilize the defect,
one can constrain (or fix) the degrees of freedom of a number of atoms
surrounding the defect of interest \cite{Boni2}.
These constraints on the configurational space, if judiciously implemented, usually do not
prevent the study of the physical properties of the (defected) system;
as an example, the binding energy of a $^3$He atom to dislocation cores turned out to be
in agreement with experimental data \cite{Corb}.
A distinct issue is the investigation of the nature (defected or not)
of the ground state of the extended system; in this case, such strategies
are not straightforwardly convincing and some doubt remains on the level of confidence to give
to the results:
beside the finite size effects, one also has to check to what extent the results are
influenced by the choice of boundary conditions. In fact,
the constraint imposed by the PBC on the configurational sampling of a crystalline solid is
both stronger and subtler than what is usually believed, due to commensurability effects between
crystal unit cell and SB \cite{Swop}.

In the present paper we address two topics. The first is to study if it is true that vacancies are not a
viable route to supersolidity because multiple vacancies coalesce and lead to phase separation \cite{Fate}.
The second topic is to study what we learn from quantum
simulations by studying large systems {\it without} using PBC.
On the first topic we review and extend
a systematic study of many vacancies (up to 30) in two-dimensional (2D) 
solid $^4$He based on exact $T=0$ K quantum simulations.
A pure 2D solid is of interest as a simple model for out of registry
solid $^4$He adsorbed on planar substrates, and
new experimental investigations \cite{Fuku} are
under way to answer the question of whether a supersolid
state is present also in adsorbed $^4$He.
In no case do we find that the crystalline order is unstable, at least as long as the
concentration of vacancies is below 2.5\%.
Multiple vacancies were found highly correlated with a tendency to form 
linear structures. Actually, when $n_v>10$ vacancies transform themselves into dislocations.
This resembles the behavior of multiple vacancies in classical solids at low temperature \cite{Lech}.
In our quantum crystal such dislocations are very mobile,
allowing exchange of particles across the system;
in fact, we find ODLRO in the one-body density matrix $\rho_1$ computed for the finite
simulated system. In presence of disorder, this gives the possibility of establishing 
a well defined phase \cite{Ander}, at least locally, in the extended system.
Should the exact ground state of $^4$He in 2D not be an ideal perfect crystal, our results
imply that the zero--point disorder does not consist of vacancies but of dislocations
and that the extended system could be supersolid.

As second topic we study systems {\it free} of PBC by confining the $^4$He atoms
in a finite region of space by an external potential.
The main aim is to investigate the true nature of the ground state, i.e.
to determine if defects restrict themselves in the interfacial region near to 
the confining potential, triggered by the mismatch between the confining geometry and the 
triangular lattice of 2D solid $^4$He, or if defects also permeate the inner region.
We give evidence that once the equilibrium is reached, the defects, even if initially placed 
in the center of the system, essentially restrict themselves in the interfacial region, leaving the
inner region a regular triangular lattice.
Moreover, we find that the inner crystal is compatible with the ideal crystal simulated in a
periodically repeated box (i.e. in standard simulations with PBC).
Given the size of the studied systems, this gives an upper bound on the concentration of
any ground state defects, which must be below $3 \times 10^{-3}$.

The paper is organized as follows:
Sec.~\ref{sec:meth} deals with the {\it exact} $T=0$ K shadow path integral ground state (SPIGS) method
and the definition of quantities used in measuring the crystal disorder.
Details on the simulations and our results on the periodically repeated crystal are presented in Sec.~\ref{sec:deta}.
Sec.~\ref{sec:zpd} contains a discussion on the issue of zero--point defects and on
the influence of using PBC.
Sec.~\ref{sec:resu} contains details on the simulations and results on the confined 2D crystal. 
Conclusions are given in Sec.~\ref{sec:conc}.

\section{Simulation details}
\label{sec:meth}

{\it The SPIGS method:} We study solid $^4$He via {\it exact} simulation methods based on
the Shadow Path Integral Ground State (SPIGS) \cite{spig1,spig2} method.
SPIGS is an extension of the Path Integral Ground State (PIGS) \cite{pigs} method.
The aim of PIGS is to improve a variationally optimized trial wave function $\psi_T$ by
constructing, in the Hilbert space of the system, a path which connects the given $\psi_T$ with the exact
lowest energy wave function of the system, $\psi_0$, constrained by the choice of the number of particle 
$N$, the geometry of the SB, the boundary conditions and the density $\rho$.
During this ``path'', the correct correlations among the particles arise through the ``imaginary
time evolution operator'' $e^{-\tau\hat H}$, where $\hat H$ is the Hamiltonian operator.
For a large enough $\tau$, an accurate representation for the lowest energy state wave function
is given by $\psi_\tau=e^{-\tau\hat H}\psi_T$, which can be written analytically by discretizing
the path in imaginary time;
this maps the quantum system into a classical system of open polymers \cite{pigs}.
An appealing feature peculiar to the PIGS method is that, in $\psi_\tau$, the variational ansatz
acts only as a starting point, while the full path in imaginary time is governed by
$e^{-\tau\hat H}$, which depends only on the Hamiltonian operator.
We have recently shown that $\psi_\tau$ does not need to be a variational optimized wave function:
PIGS results for large enough $\tau$ are unaffected by the choice of
$\psi_T$ both in the liquid and in the solid phase \cite{Vita,Ross3} thus providing an unbiased
exact $T=0$ K QMC method.
Within SPIGS a shadow wave function (SWF) \cite{Viti,Moro} is taken as $\psi_T$ and this choice
was shown to greatly accelerate convergence to $\psi_0$.
Another feature of the SPIGS method is that it recovers the solid phase via a
spontaneously broken translational symmetry \cite{spig1,spig2}, like in Path Integral Monte Carlo (PIMC),
and there is no constriction on the
atomic positions, so it is particularly useful in studying crystals with defects.

{\it Measuring crystal order and fluctuations:}
We check the presence of crystalline order by monitoring the static structure factor for the
presence of Bragg peaks and, via a Delaunay triangulation \cite{Fate} (DT) of the sampled 
configurations, we compute the particle coordination number in order to estimate the amount of 
local disorder in the system \cite{note}.
In an ideal perfect 2D triangular crystal, with the atoms in their equilibrium positions, each atom 
is linked to 6 other atoms in the DT.
Atoms with coordination number not equal to 6 are then a measure of fluctuations and of
local disorder in the crystal.
In periodically repeated systems in 2D one proves a conservation law for the coordination 
numbers:
\begin{equation}
 \label{conslaw}
 \sum_{i=3}^5(6-i)N_i=\sum_{i+7}^\infty (i-6)N_i
\end{equation}
where $N_i$ is the number of $i-$coordinated atoms, so that we can consider only $N_i$ with
$i<6$.
Since it is always verified that $N_3=0$, the quantity $\tilde X_d=2N_4+N_5$ is usually taken as
an estimate of local disorder in the system \cite{Ches2}.
However, even in an ideal (defect-free) 2D quantum crystal the coordination is 6 only on the 
average:
atoms are not always 6-fold coordinated due to their large zero--point motion.
Then, a more useful quantity measuring the net amount of disorder in a crystal with $n_v$ vacancies,
for instance, is the difference between the observed  $\tilde X_d$ and that of the corresponding ideal
perfect crystal:
\begin{equation}
 X_d=(2N_4+N_5)_{n_v}-(2N_4+N_5)_{n_v=0}.
\end{equation}

We have studied also the orientational order parameter $\Psi_6$ to quantitatively characterize
the quality of crystalline structures.
The local order parameter that measures the degree of 6-fold-orientational ordering is defined \cite{Frenk} as
\begin{equation}
 \label{psi6}
 \psi_6(\vec{r}_k) = {1 \over n(k)} \sum_{j=1}^{n(k)} e^{i6\theta_{kj}}
\end{equation}
where $\theta_{kj}$ is the angle between the vector $\vec r_k-\vec r_j$ and a fixed direction in the 
plane (chosen here to be the positive $x$ direction) and the sum over $j$ extends over all $n(k)$ 
nearest-neighbor of the $k$-th atom as provided by the DT. 
The global order parameter associated with bond-orientational order is then obtained as an 
average over all particles
\begin{equation}
 \label{Psi6}
 \Psi_6 = {1 \over N} \sum_{k=1}^N \psi_6(\vec{r}_k)
\end{equation}
In a static ordered triangular solid, we have $n(k) = 6$ and $\theta_{kj}$ is a multiple
of $\pi/3$ for all $j = 1,\dots,6$.
In such a case, $\mid \langle \Psi_6 \rangle \mid = 1$ and the plot of $\langle \Psi_6 \rangle$ 
on a complex plane will result in a single spot on a circumference with unit radius.
Because of the fluctuations due to the zero--point motion we expect that, even in the perfect
crystal, this spot will be broadened along the circumference and toward the center.
In the presence of vacancies or other defects we expect significantly less orientational order and the 
plot of $\Psi_6$ should extend over a larger portion of the circle (in the limiting case of the 
liquid system, it turns in a distribution centered around the origin).

\section{Results: periodically replicated crystal}
\label{sec:deta}

Recently we have performed the first systematic study of multiple vacancies in two dimensional (2D) solid
$^4$He \cite{Fate}.
Dealing with low temperature properties, $^4$He atoms are described as structureless zero--spin
Bosons interacting through the standard HFDHE2 Aziz potential \cite{Aziz}.
From the computed equation of state we found the freezing and the melting densities to
be, respectively, $\rho_f=0.0672$ \AA$^{-2}$ and $\rho_m=0.0724$ \AA$^{-2}$.
In our SPIGS computation for the periodically repeated system we have used the pair-product
approximation \cite{Cepe} for the imaginary time projector;
as a compromise between good accuracy and reasonable computational effort
$\delta\tau$ has been set to $1/40$ K$^{-1}$.
A total projection imaginary time $\tau = 0.775$ K$^{-1}$ ensures convergence to the ground state.
We have simulated a 2D crystal at densities slightly above the melting
density.
The geometry of the SB is such that it is compatible with a regular triangular lattice and the study
has been performed with $M=240$, 480, 960 and 1440 {\it lattice positions} and PBC were 
applied in all directions.
When $N=M$ we have an ideal ordered crystal.
By removing atoms from the initial ordered configuration we can study a system with a variable $n_v = M-N$
number of putative vacancies.
$n_v$ ranged from 0 to 30, with a maximum vacancy concentration $x_v=n_v/N$ of 0.025.
For larger values of $x_v$ the crystalline order is lost.
After removing $n_v$ particles from the starting ideal ordered configuration the dimensions of the SB
were rescaled to restore the original crystal density $\rho$.
As a test of convergence we performed simulations starting from different positions of the $n_v$
vacancies.

\begin{figure*}
 \includegraphics[width=0.75\textwidth]{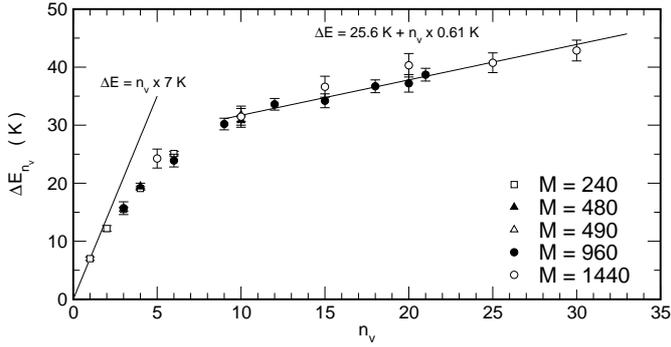}
 \caption{\label{f:att} Defect formation energy $\Delta E_{n_v}$ at constant density as a function of $n_v$
          in 2D solid $^4$He at $\rho = 0.0765$\AA$^{-2}$ computed in boxes with different
          lattice site numbers $M$.
          Dashed lines are linear fit to the data.}
\end{figure*}

\begin{figure*}
 \psfrag{Im}[]{\begin{sideways}\footnotesize{$\Im[\langle\Psi_6\rangle]$}\end{sideways}}
 \psfrag{Re}[]{\footnotesize{$\Re[\langle\Psi_6\rangle]$}}
 \begin{tabular}{lll}
  \hspace{-0.5cm}
  \includegraphics*[width=0.45\textwidth]{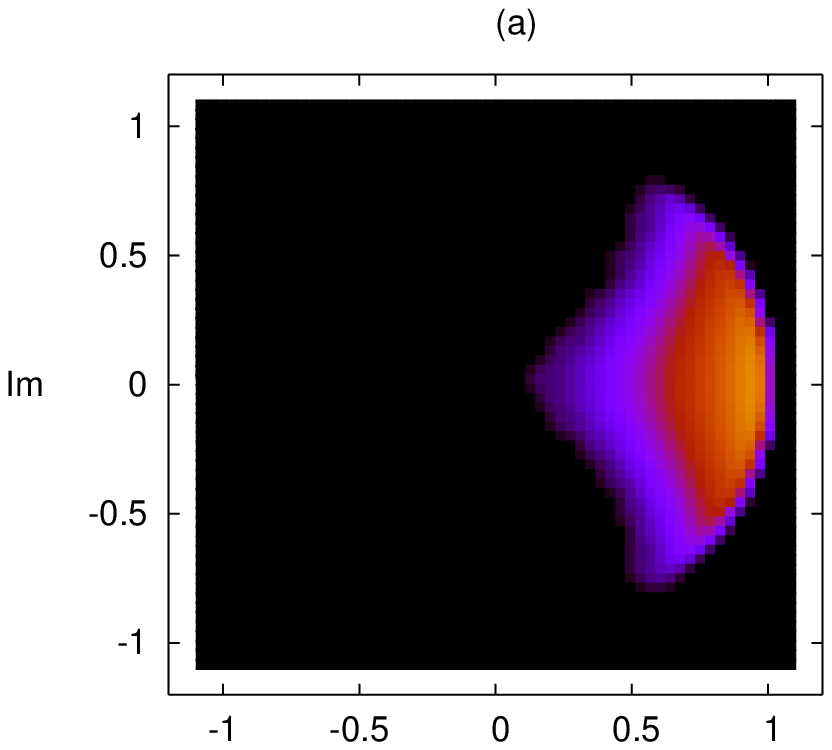} &
  \hspace{-2.4cm}
  \includegraphics*[width=0.45\textwidth]{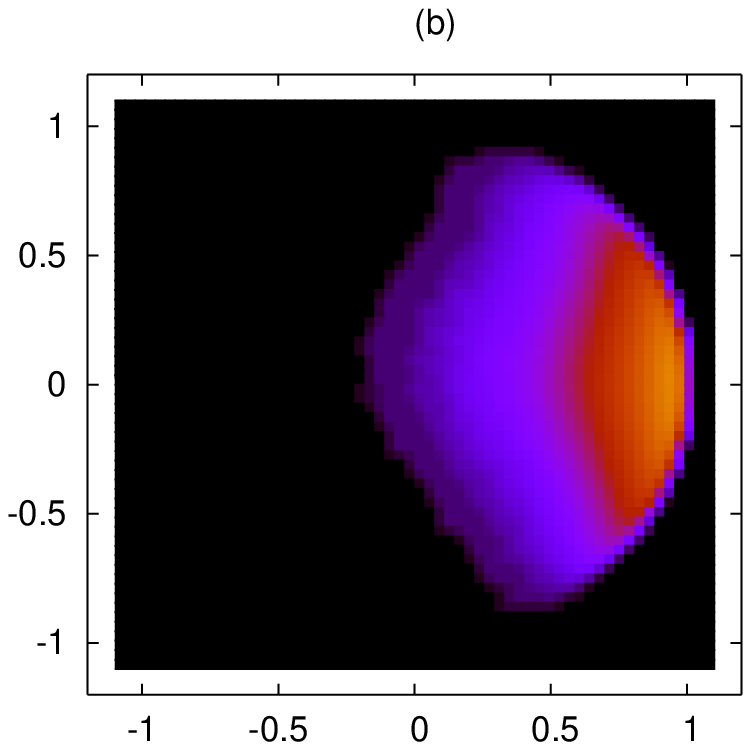} &
  \hspace{-2.4cm}
  \includegraphics*[width=0.45\textwidth]{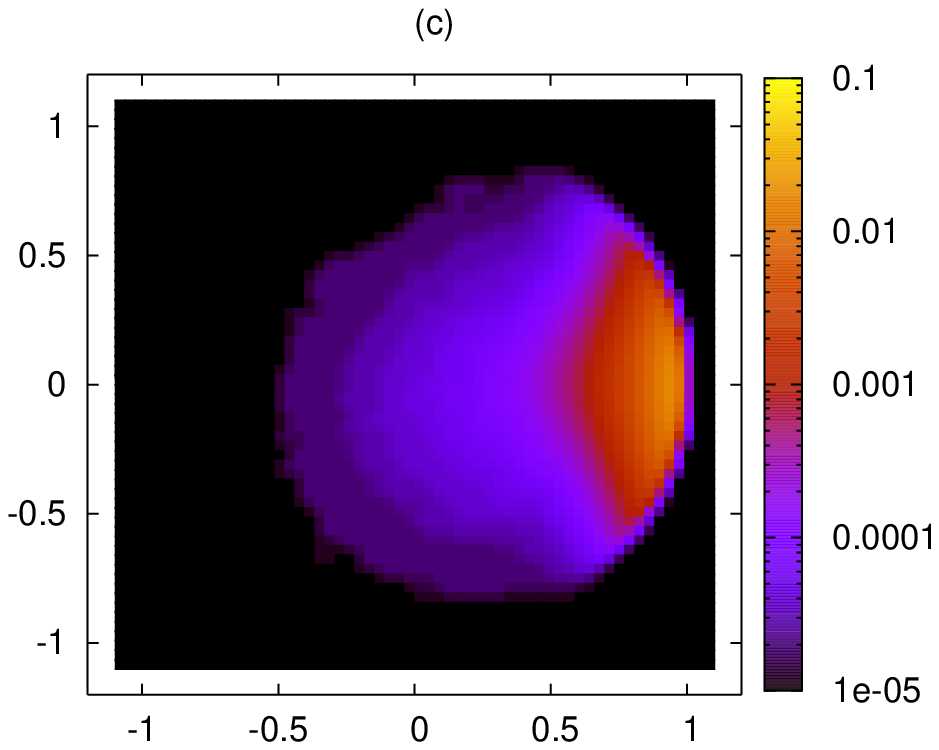}\\[-2.0cm]
  \hspace{-0.5cm}
  \includegraphics*[width=0.45\textwidth]{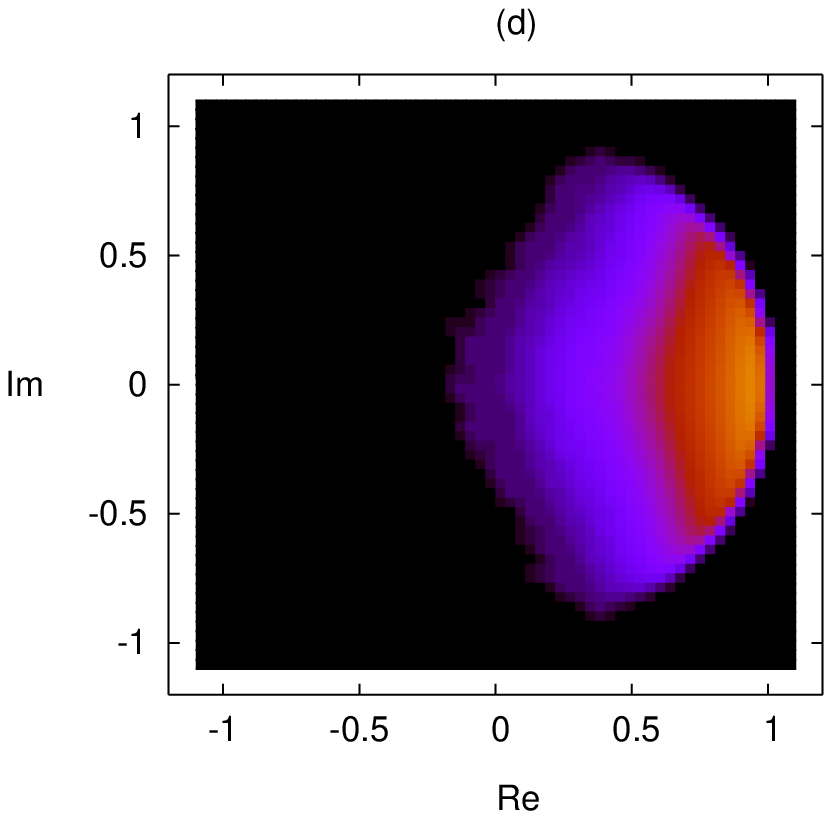} &
  \hspace{-2.4cm}
  \includegraphics*[width=0.45\textwidth]{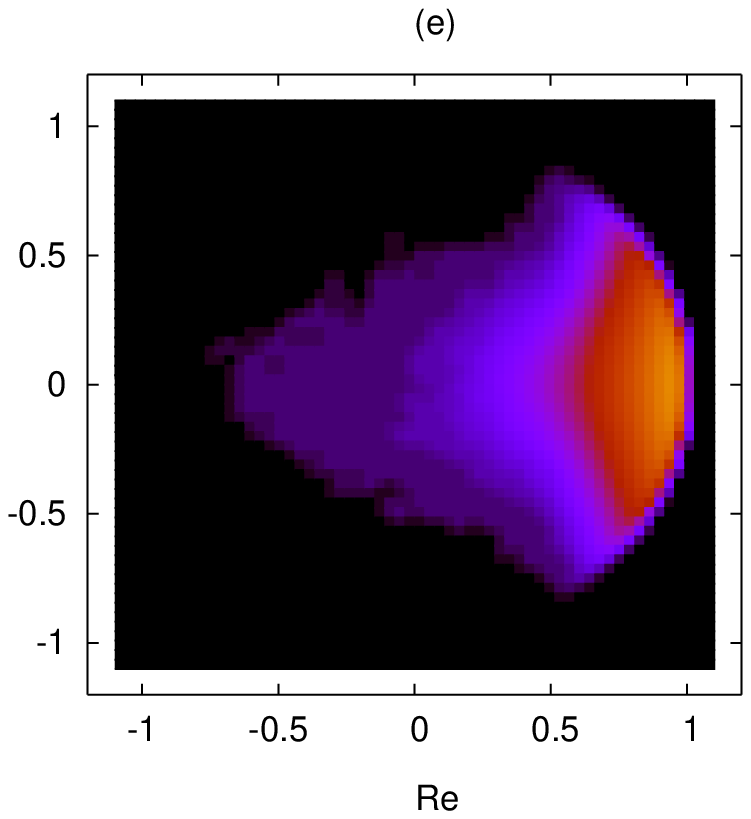} &
  \hspace{-2.4cm}
  \includegraphics*[width=0.45\textwidth]{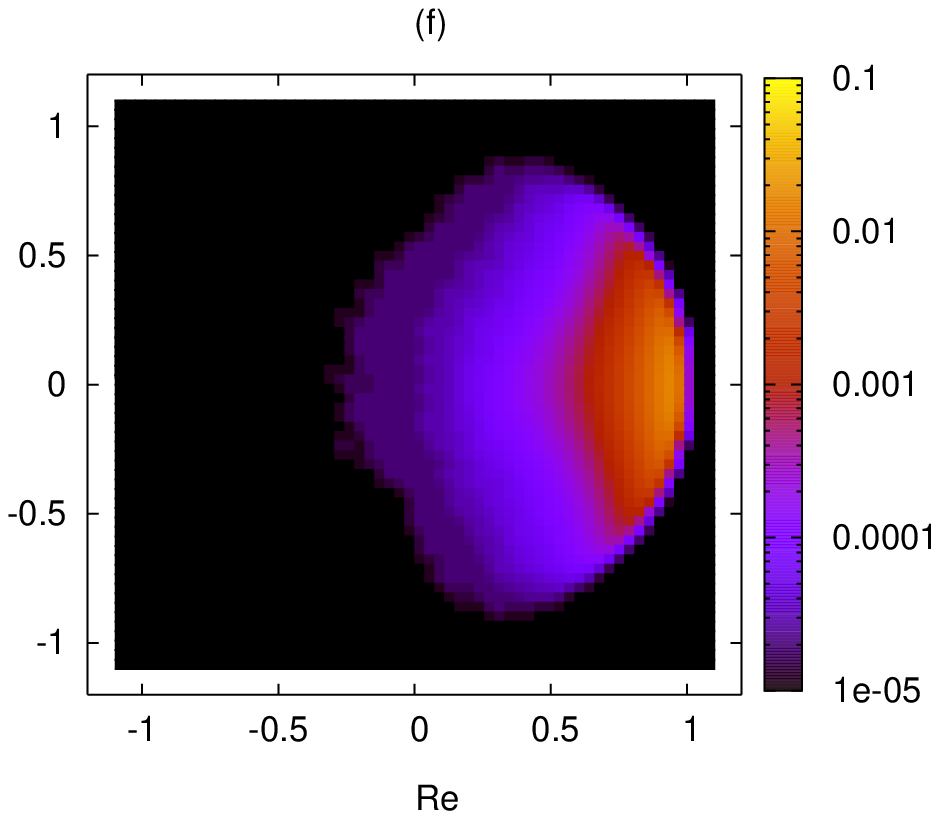}
 \end{tabular}
 \caption{(Color online) Intensity plots of the probability of $\Psi_6$ on the complex plane
         for a bulk 2D $^4$He crystal at $\rho=0.079$ \AA$^{-2}$, for the (a) ideal perfect crystal
         with $N=572$, (b) crystal with a vacancy ($N=571$), (c) crystal with 4 vacancies ($N=568$)
         (d) crystal with dislocations ($n_v=10$) with $N=562$, (e) crystal with an interstitial 
         ($N=573$), (f) crystal with 4 interstitials (dislocations) with $N=576$. 
          Note the logarithmic intensity scale.}
 \label{fig3}
\end{figure*}

The crystalline order is stable for all the cases considered and the properties
of the system display a dependence on $n_v$ rather than on $x_v$.
Both the diagonal and the off-diagonal properties were found to exhibit a double regime behavior\cite{Fate}.
For $n_v<6$ the defects are easily identified in the snapshots of the atomic positions and
the defect formation energy $\Delta E_{n_v}$ is found to be monotonic and systematically sublinear
with $n_v$. This is consistent with the presence of an attractive interaction among vacancies
(as confirmed by the vacancy-vacancy correlation function \cite{Fate}).
For $n_v \gtrsim 10$ $\Delta E_{n_v}$ becomes linear in $n_v$, with a cost of 0.61K for each
additional vacancy, i.e. the formation energy for an additional vacancy in this regime costs about ten
times less than the energy (7 K) required for a single vacancy (see Fig.~\ref{f:att}).
A similar double regime behavior has been found also in the static structure factor and in the
amount of disorder $X_d$.
For $n_v<6$ the main Bragg's peaks are found as sharp as in the perfect crystal but with a
decreased height and $X_d$ is found to increase with $n_v$ (as expected for an increasing disorder).
When $n_v\gtrsim10$ the behavior is quite different: the main Bragg's peaks are slightly broadened 
but with an integrated intensity that is almost independent of $n_v$ and also $X_d$ does not depend
on $n_v$.
By looking to the sampled configurations we realized that multiple vacancies for $n_v\lesssim6$
are highly correlated with a preference to form fluctuating linear structures.
On the other hand when $n_v\gtrsim10$ the vacancies inserted in the initial configuration lose 
their identity and transform themselves into quantum dislocations \cite{Fate}.
In the range 6--10 there is a crossover between the two regimes.
We have now analyzed such multiple vacancy systems in terms of the orientational order parameter 
$\Psi_6$.
$\Psi_6$ also displays a similar double regime behavior.
Some examples of the obtained results are reported in Fig.~\ref{fig3}.
By comparing the result of $\langle\Psi_6\rangle$ in the ideal perfect crystal (Fig.~\ref{fig3}a) 
with the results for the crystal with 1 vacancy (Fig.~\ref{fig3}b), 4 vacancies (Fig.~\ref{fig3}c) 
and 10 putative vacancies (which equilibrate into dislocations, Fig.~\ref{fig3}d), one can see that, in the cases
when disorder is present, $\langle\Psi_6\rangle$ fills more of the complex plane far from
the point $\Re[\langle\Psi_6\rangle]$=1 on the real axis with respect to the perfect crystal case.
However, in going from the 4 vacancies to the dislocations case, the signal of disorder decreases,
inverting the trend from 1 to few vacancies; again this is a signature of the double regime
behavior found in monitoring other quantities.
An interstitial (Fig.~\ref{fig3}e) gives a distinct signature of disorder with respect to that of a 
vacancy; 4 interstitials give a disorder map in $\langle\Psi_6\rangle$ that is less extended than that for
a single interstitial, showing some cooperative correlations also among interstitials.
In fact, by looking at the DT of the sampled configurations we can recognize the presence of
dislocations also in this case.

\begin{figure*}
 \begin{tabular}{lll}
  \hspace{-0.5cm}
  \includegraphics*[width=0.33\textwidth]{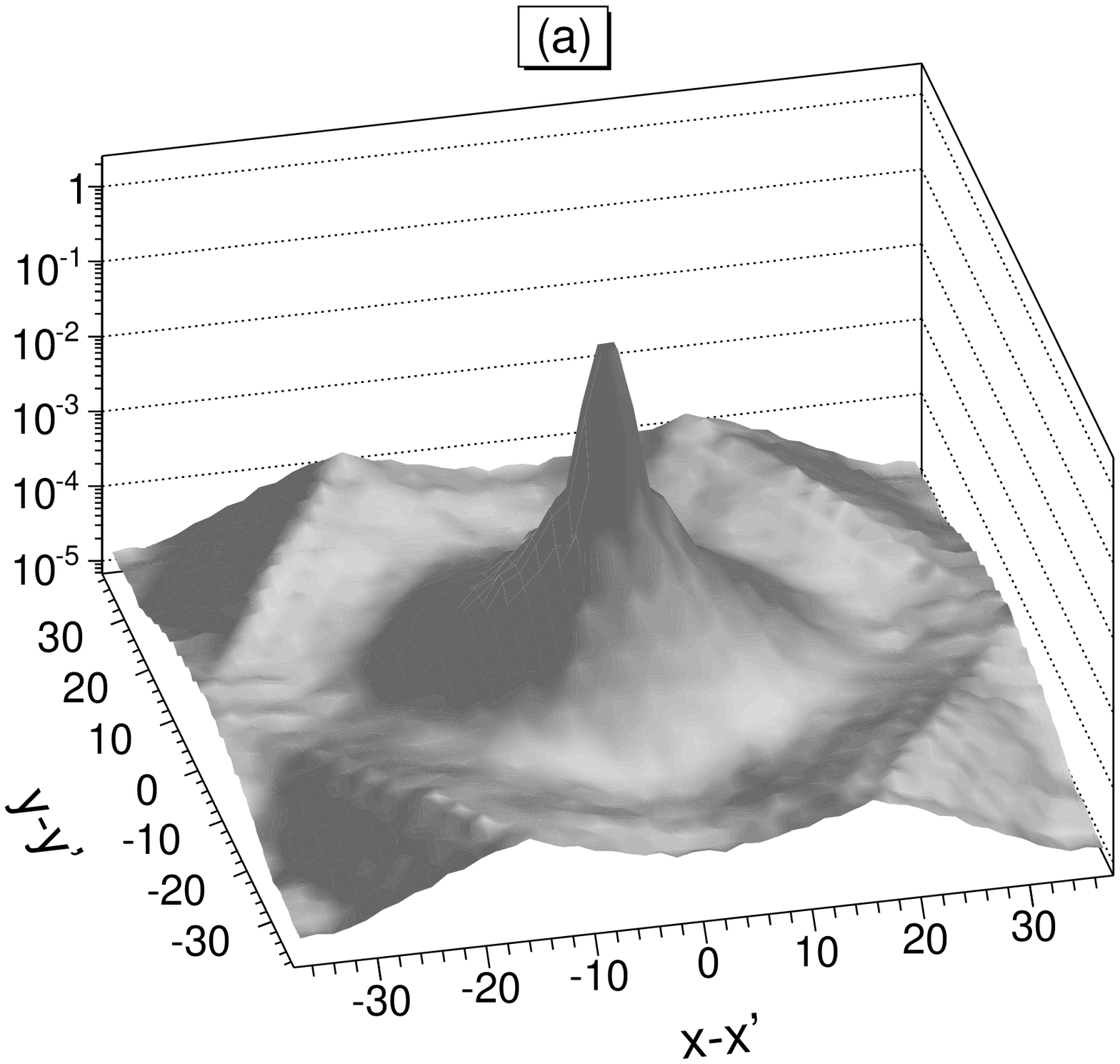} &
  \hspace{-0.5cm}
  \includegraphics*[width=0.33\textwidth]{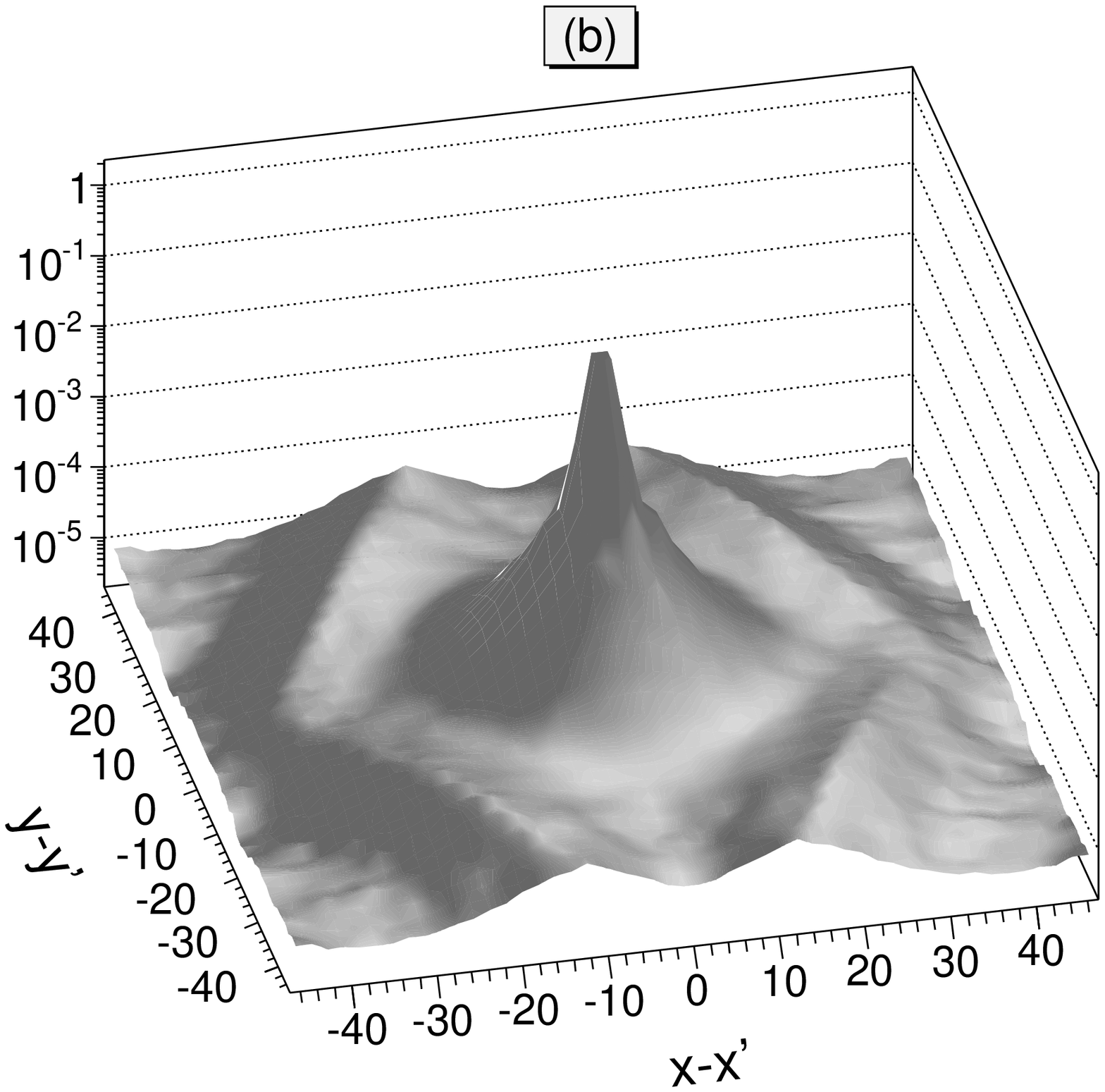} &
  \hspace{-0.5cm}
  \includegraphics*[width=0.33\textwidth]{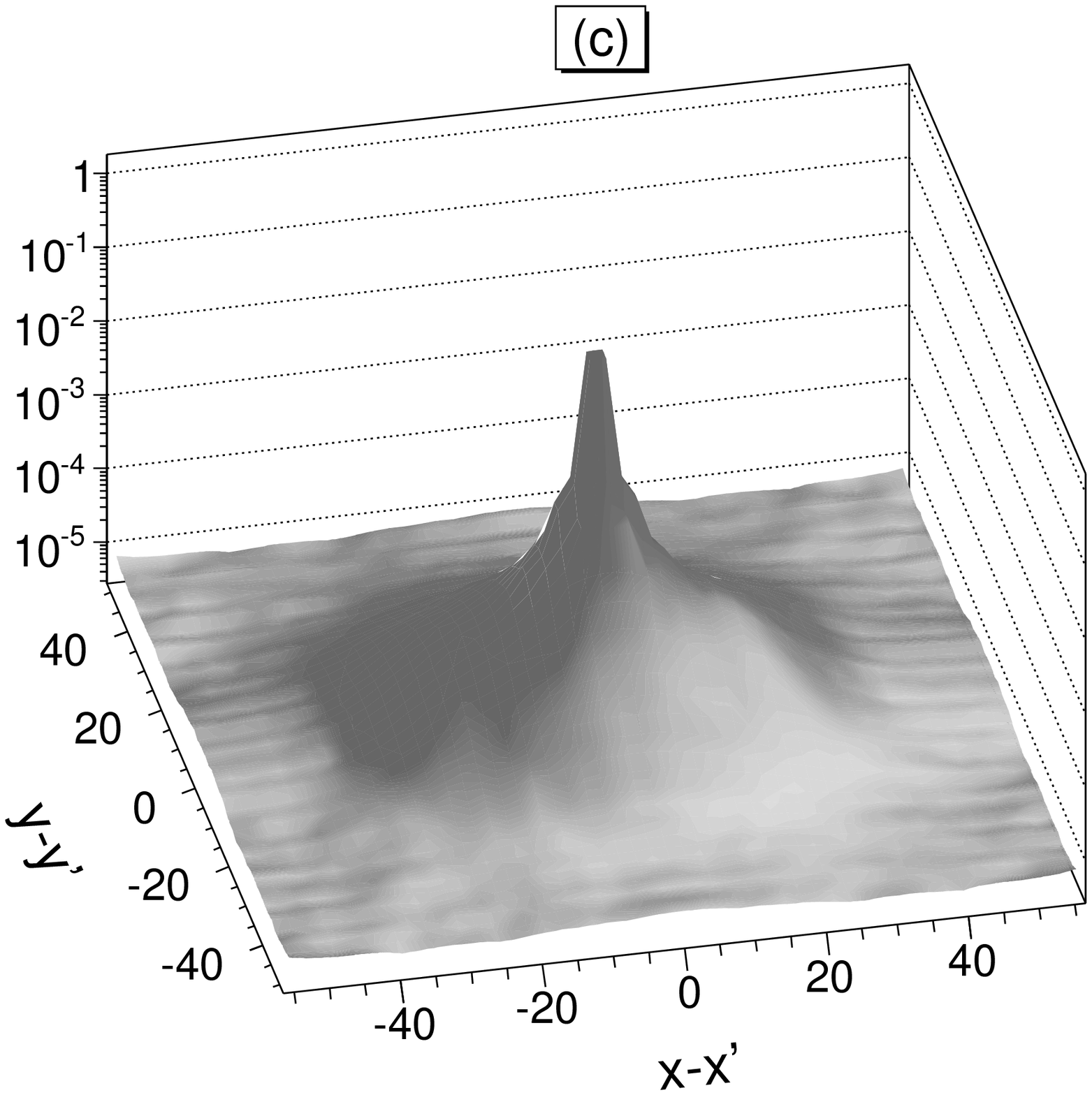}
 \end{tabular}
 \caption{One--body density matrix $\rho_1(\vec{r},\vec{r}\,')$
          in the $x-x',y-y'$ plane computed in a defected $^4$He 2D crystal ($n_v=10$) at
          $\rho = 0.0765$\AA$^{-2}$ for different numbers of lattice sites $M$:
          (a)$M=480$ (b) $M=700$ (c) $M=960$ .}
 \label{odlro}
\end{figure*}

This double regime is found also in the large distance behavior of the one-body density matrix
$\rho_1$.
For small $n_v$, as $n_v$ increases, the plateau in the $\rho_1$ tail, which is the signature of ODLRO,
decreases, and this can be interpreted as an effect of the vacancy--vacancy interaction.
Should this trend continue into the large $n_v$ regime, no ODLRO could be present
in the thermodynamic limit, implying that an extended crystal with vacancies would not be supersolid.
However we find a different behavior:
for $n_v \gtrsim 10$ the plateau in the large distance tail of $\rho_1$ is restored 
(see Fig.~\ref{odlro}) mainly due to the ability of the dislocation cores to transfer particles 
among them.
In fact, dislocations turn out to be very mobile and are able to induce exchanges of particles across 
the whole system, which is a necessary condition for ODLRO.
Unfortunately, no size scaling analysis on $\rho_1$ has been done yet in the dislocation
regime because of the prohibitive computational cost of off-diagonal simulations of systems large
enough to accommodate a large number of dislocations.
In Fig.~\ref{odlro}a and Fig.~\ref{odlro}b one sees the presence of ridges in the tail of $\rho_1$.
As discussed in detail in Ref.~\cite{Fate}, these ridges are due to commensuration effects
between dislocations and SB size.

An important question is the behavior of many vacancies in a 3D crystal.
Based on finite temperature simulations, statements in the literature \cite{Boni} are found that in
presence of many vacancies the crystal becomes unstable against separation into a vacancy--rich
and a crystalline vacancy--free phase.
We found no such instability at $T=0$ K not only in 2D but also in 3D 
for as many as 98 vacancies as discussed in Ref.~\cite{Fate}.

\section{Periodic boundary conditions, the crystalline state and the issue of zero--point defects in solid $^4$He}
\label{sec:zpd}

All simulations deal with a finite number $N$ of particles and, in order that such finite
size systems mimic a bulk system, PBC turn out to be very useful.
In a fluid system the extrapolation to the bulk limit of the simulation results for short range
properties is rather straightforward, unless one is close to a critical point, because the
results depend smoothly on $N$.
The situation is more complex for a crystalline solid because simulated properties are affected
also by commensuration effects between the SB and the crystalline unit cell.
As an example of this consider the following.
The stable phase of solid $^4$He at low $T$ and pressure is the hcp crystalline state.
However, from exact quantum simulations of realistic models of $^4$He it turns out that all
three phases hcp, fcc and bcc are stable (with a different value of the energy) for a 
suitable choice of the SB geometry and of $N$ \cite{spig3}; i.e. the average density, $\rho$, has to be
large enough and the sides of the SB have to be integer multiples of the sides of the unit
cell of the crystal under study.
Finite size scaling for a given phase is performed by considering a special set of numbers of
particles, for instance 180, 448, 900 for hcp, but this explores a special set of states predetermined
by the choice of SB and $N$.
The conclusion is that finite size scaling is very special in a crystalline solid and it does
not give an unbiased estimate of the properties of the bulk system.
One might think that a way out of such a limitation is to start from a disordered configuration 
and let the system crystallize into the preferred state.
The problem is that PBC again constrain the system.
In fact, compatibility between a crystal lattice and PBC for a given SB is present only for certain
orientations of the crystal axis with respect to the axis of the SB.
The spontaneous crystallization of the particles has essentially zero probability to pick up such
special orientations and, in fact, computer experiments \cite{Pede2} show that
crystallization does indeed take place, but the crystal axis take arbitrary orientations with
respect to the SB so that the crystalline order is deformed in order to comply with the PBC.    

For similar reasons one cannot expect to see during the computation the spontaneous appearance of
a vacancy.
An objection \cite{Boni,n0T2,Cepe2} to the presence of ground state vacancies is that such defects cost large energies,
quantum simulations \cite{Pede,spig1,Boni,Cepe2} agree on a value of order of 15 K in 3D at the 
melting density, so that no defect should be present at low $T$.
This, however, fails to recognize that those simulations addressed specifically the question
of the energy cost of a vacancy as an excited state and this is a separate question from the 
presence of vacancies in the ground state.
It is instructive to consider the lattice gas model introduced by Mullin \cite{Mull}.
This model displays a liquid phase, a normal solid and a supersolid phase.
In the supersolid state the occupation probability of a lattice site is less than unity
and delocalized vacancies are present (more precisely vacancy-interstitial pairs)
as an effect of zero point motion. Yet, the excitation spectrum has a gaped vacancy-interstitial branch.
Anderson \cite{Ande2} has argued that in a Bose quantum crystal a unit population of
a lattice site is incompatible with quantum mechanics and that some vacancies have to be present
in a highly quantum system like solid $^4$He.

One way to address the question of the presence of vacancies in the ground state of solid $^4$He
is to proceed in a way similar to what is done in classical statistical mechanics in computing the
equilibrium concentration of vacancies in a classical solid.
The method is based on considerations of a macroscopic system, not the one that is simulated, and
a basic assumption is that vacancies can be treated as weakly interacting objects.
Within variational theory,
using a SWF as ground state, a finite concentration $x_v$ of zero--point vacancies has been found with 
$x_v\simeq1.4\times10^{-3}$  at melting density \cite{Ross}.
This is a sizeable concentration not too far from the actual experimental upper bound.
Extension of this computation to the case of the exact ground state is a difficult and open 
problem also because one has to take into account that vacancies are strongly interacting and
tend to transform themselves into dislocations, at least in 2D, as discussed in the previous section.

Another approach to study the presence of ground state vacancies is to get away from PBC by
considering a confined system.
This is the topic of the next section. 
 
\section{Confined crystal}
\label{sec:resu}

{\it Specific simulation details:}
In the study of confined systems we have opted for an implementation of the worm algorithm \cite{n0T1}
with a fixed number of particles and for the pair-Suzuki approximation \cite{Ross3} for the
imaginary time propagator $e^{-\delta\tau\hat H}$; $\delta\tau$ has been set to $1/360$ K$^{-1}$ 
and the total imaginary projection time $\tau=0.775$ K$^{-1}$ is equal to the value used in the
simulations of the periodically replicated crystal.

\begin{figure*}
 \includegraphics*[width=0.75\textwidth]{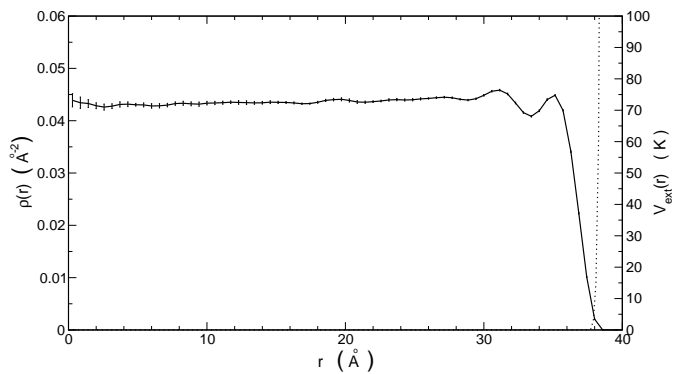}
 \caption{(solid line) Radial density profile, $\rho(r)$, for a system with $N=187$ and $R=40.0$ \AA.
         (dotted line) Confining potential, $V_{\rm ext}(r)$}
 \label{rho}
\end{figure*}

The confining potential has been chosen to be circular symmetric so that all directions are 
equivalent and, in addition, this ensures minimal mismatch with the crystalline triangular 
structure so that the interfacial region should be minimal.
Since we are not interested in simulating any specific confinement, we have chosen the confining
potential $V_{\rm ext}$ to be only repulsive and to represent an hypothetical surrounding infinite 
$^4$He bulk crystal at density $\rho_{\rm ext}$ with the atoms considered as a continuum.
Then $V_{\rm ext}$ reads
\begin{equation}
 V_{\rm ext}(r) = \left\{
 \begin{array}{lll}
  V_{\rm LJ}(R-r) - V_{\rm LJ}(R-r_{\rm min}) & \quad & R-r_{\rm min}<r<R \\
  0                                           & \quad & r < R-r_{\rm min}
 \end{array}\right.
\end{equation}
where $R$ is the radius of the confining potential and
$V_{\rm LJ}$ is obtained by integrating a standard 12-6 Lennard-Jones potential on the
half-plane:
\begin{equation}
 V_{\rm LJ}(r) = \frac{3\pi\varepsilon\sigma^6}{8}\rho_{\rm ext}\left[\left(\frac{21\sigma^6}{80r^{10}}\right)
                                                           -\left(\frac{1}{r^4}\right)\right]
\end{equation}
and $r_{\rm min}$ is the position of the minimum in $V_{\rm LJ}$.
As $\varepsilon$ and $\sigma$ we use the typical Lennard-Jones parameters for $^4$He, i.e.
10.22 K and 2.556 \AA~ respectively, and $\rho_{\rm ext}$ is set to 0.0765\AA$^{-2}$.
Such a confining potential fails in accounting for the confining curvature, but the curvature effect is small
for the large values of the radius $R$ here considered.
The initial configuration is obtained by building a regular triangular lattice at a density
$\rho_S$ and discarding all the particles falling out the disk of radius $R-r_{\rm min}$.
The number of particles inside this disk is referred to as $N_0(\rho_S)$.
If $\rho_S$ is below a certain value the atoms quickly lose the initial crystalline order and the system
evolves into a liquid state.
As an example, in Fig.~\ref{rho} the case of a confining potential with $R=40$ \AA~ and $N_0(\rho_S)=187$ 
($\rho_S = 0.0435$ \AA$^{-2}$) is shown.
From the plot of the radial density profile it is evident that the confinement gives rise to an 
interfacial region that covers an annulus whose thickness is not larger than about 8 \AA.
By increasing the number of particles at fixed radius $R$, we find that the particles become
more and more localized and beyond a certain value of $N$ the solid order becomes visually evident.
For the solid phase, such interfacial region is expected to be thicker because of the effect of 
the mismatch between the triangular geometry of the confined crystal and the circular one
of the confining potential.
We have considered different values of the radius $R$ and we find that the interfacial region
never penetrates more that 15 \AA~ inside the system.

In order to promote the presence of defects in the inner region, additional simulations are performed
by subtracting or inserting particles at random positions from
an equilibrated configuration for the system with $N_0(\rho_S)$ particles.
For the confined crystal the quantity $\tilde X_d$, introduced in the previous section,
is no longer appropriate because of the lack of a conservation law like
the one in (\ref{conslaw}). In place of $\tilde X_d$,
as an estimate of the local disorder, we use the density $\rho_{\slashed{6}}$ of the particles that 
are miscoordinated, i.e. with coordination number different from 6.

\begin{figure}
 \includegraphics*[width=0.45\textwidth]{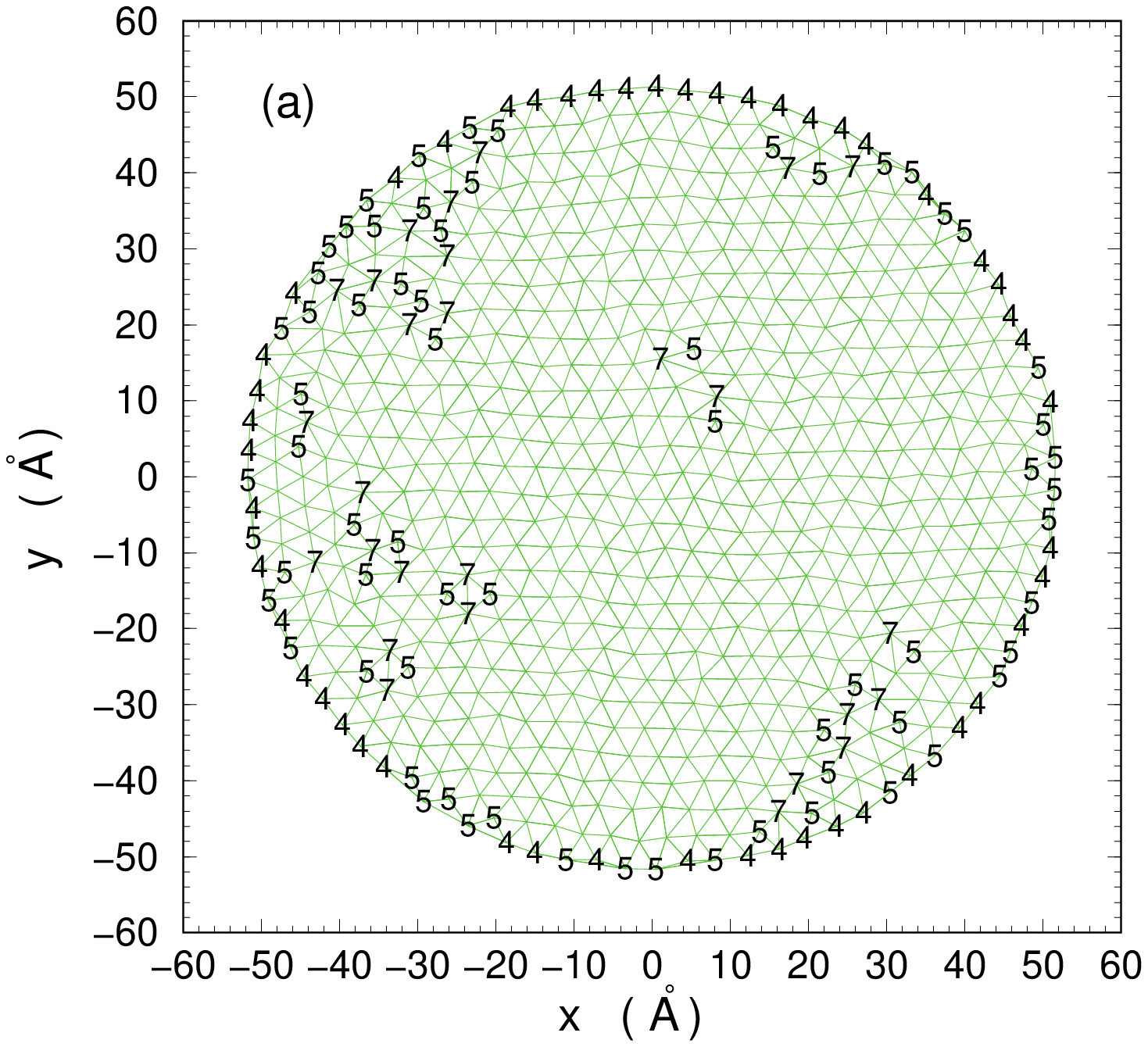}
 \includegraphics*[width=0.45\textwidth]{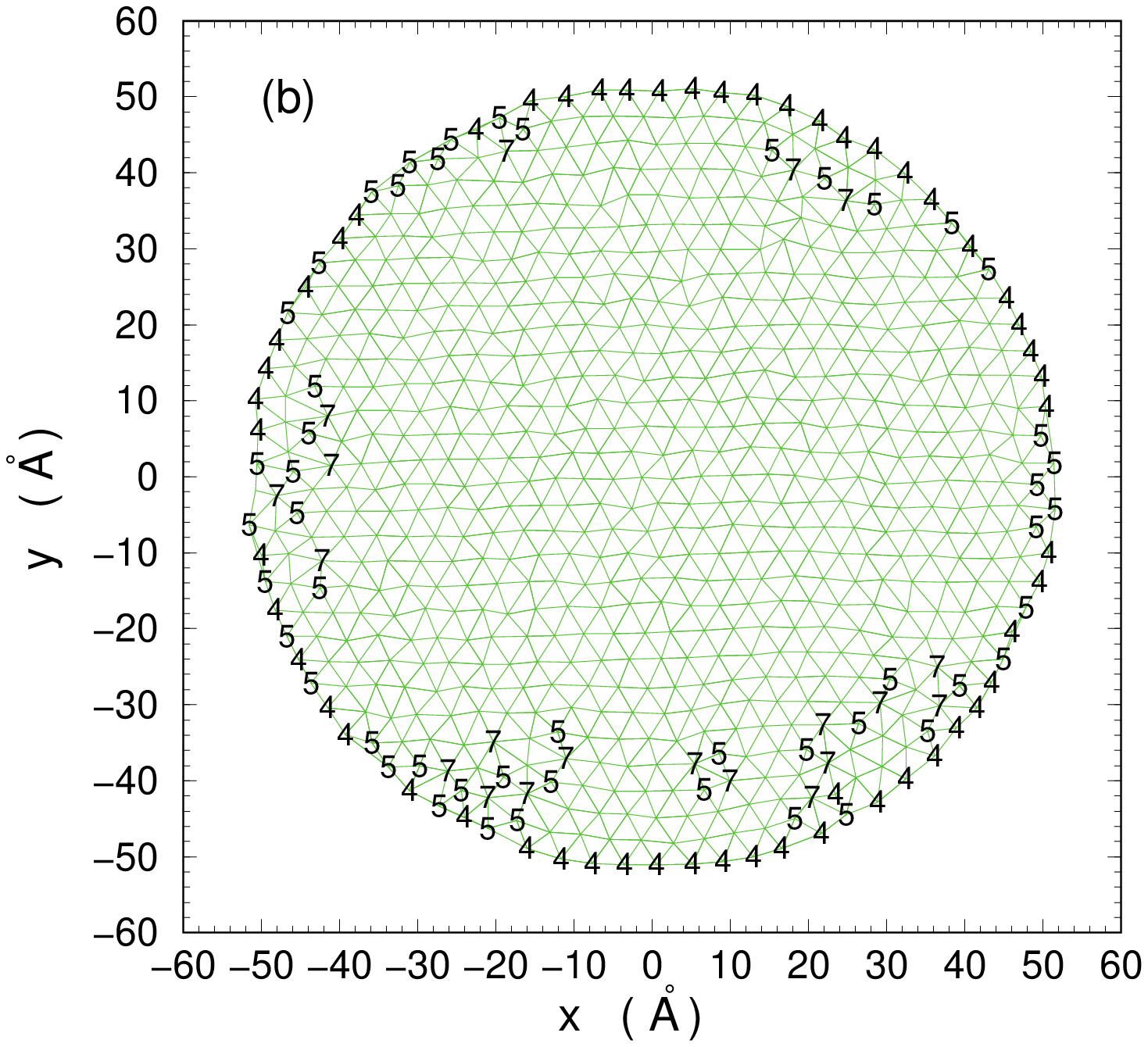}
 \caption{(color online). Delaunay triangulation of typical
          configurations of a 2D confined $^4$He crystal with
          $N=671$ and $R=54.6$ \AA~ (a) after few hundreds Monte Carlo steps from the
          random removal of 14 particles from an equilibrated configuration of the system
          with $N=N_0(\rho_S)$ and (b) when equilibrium has been reached.
          We report the coordination number for atoms only when different from 6.
          The couple 5--7 indicates a dislocation core in 2D and the clumps of two 5 and two 7
          indicate bound pairs of dislocations.}
 \label{tria}
\end{figure}

\begin{figure*}
 \includegraphics*[width=0.75\textwidth]{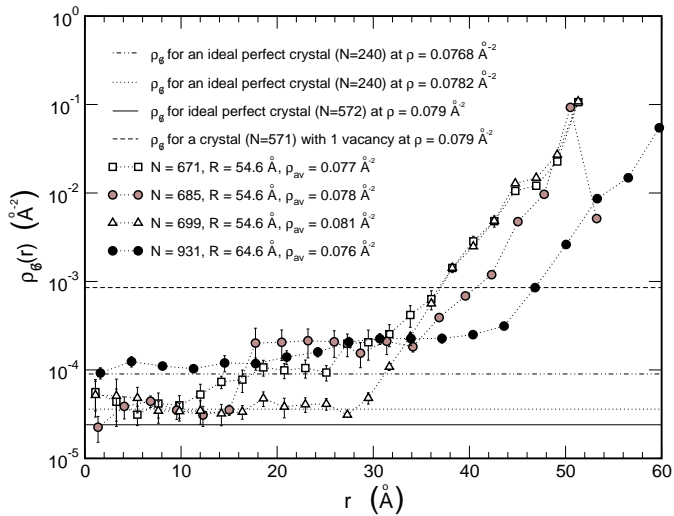}
 \caption{Local densities, $\rho_{\slashed{6}}(r)$, of particles with coordination different from 6
         as function of radial distance for a number of 2D confined $^4$He crystals. 
         For comparison with the values of bulk
         2D $^4$He crystals at similar densities are also shown (see legend).}
 \label{fig1}
\end{figure*}

{\it Results:} We have considered three different values for $R$, 44.6 \AA, 54.6 \AA~ and 64.6 \AA, that
give rise, respectively, to $N_0(\rho_S)=433$, 685 and 931 $^4$He atoms when the starting 
configuration is obtained by cutting from an ideal triangular lattice at 
$\rho_S=0.0765$ \AA$^{-2}$.
When equilibrated, the system is solid with an average density $\rho_{\rm av}$ slightly 
different from $\rho_S$ because of relaxations of the atoms in the confining potential 
($\rho_{\rm av}$ are reported in the legend of Fig.~\ref{fig1}).
In the present paper we discuss mainly the case of the intermediate radius value 
$R = 54.6$ \AA.
In addition to the ``magic'' number $N_0(\rho_S)=685$ we have considered also an initially  ``defected'' crystal 
by changing the number of particles.
From an equilibrated configuration with 685 $^4$He atoms we have removed 14 particles 
at random positions obtaining a system with $N = 671$ particles, and we have also randomly 
inserted 14 particles in order to obtain a system with $N = 699$ particles.
In the case of such initial defected state the MC evolution has two stages.
At first, after a few hundred MC steps, the crystal relaxes around the injected defects and 
we have a confined solid with defects (vacancies or interstitials) that are present also in 
the inner region.
In Fig.~\ref{tria}~(a) the DT of one typical configuration in this initial evolution for the 
$N = 671$ system is shown.
At a later stage of the MC evolution, the system reaches an equilibrium state with no sign of 
defects recorded in the inner region (see Fig.~\ref{tria}~(b)).
This is confirmed also by the density of miscoordinated particles $\rho_{\slashed{6}}$.
From the plot in Fig.~\ref{fig1} it is easy to recognize in the system with $R = 54.6$ and 64.6
\AA~ a disk of about 30 \AA~ radius in which $\rho_{\slashed{6}}$ is essentially
compatible with the expected value in periodically repeated ideal crystals at similar densities.
In Fig.\ref{fig1} $\rho_{\slashed{6}}(r)$ for a periodically repeated crystal with one vacancy is also shown.
Its value is about 40 times larger of that of the system without a vacancy.
It is clear that the values of $\rho_{\slashed{6}}(r)$ in the inner region of the confined system
are incompatible with the presence of such defect.

The suggestion of a defect-free inner region is supported also by the following argument.
Consider a periodically repeated crystal.
As discussed in Sec.~\ref{sec:meth}, even in the ideal perfect crystal there is a finite probability
to find miscoordinated particles due to the zero--point motion.
Since this is a fluctuation effect, we might expect the permanence time $t_{\slashed{6}}$, i.e. the
mean number of MCS in which a given ill coordinated particle remains ill coordinated, is smaller in an
ideal perfect crystal compared with the case of a crystal with a vacancy, because the
vacancy is a permanent defect whose dynamics is governed by the rate of vacancy jump.
Indeed, at density $\rho=0.77$ \AA$^{-2}$ we find $t_{\slashed{6}} = 4.0$ MCS for the periodically repeated
ideal perfect crystal and $t_{\slashed{6}} = 9.5$ when a vacancy is present.
A similar computation for the confined system and considering only particles with 
$r<30$ \AA~ gives $t_{\slashed{6}} = 3.0$, $2.7$ and $2.8$ in the systems with $N=671$, 685 and 699,
respectively.
Thus not only is the inner region free of defects, but it turns out to be somehow less fluctuating
than the system with PBC, presumably due to the stabilizing effect of the confining potential.

\begin{figure*}
 \psfrag{Im}[]{\begin{sideways}\footnotesize{$\Im[\langle\Psi_6\rangle]$}\end{sideways}}
 \psfrag{Re}[]{\footnotesize{$\Re[\langle\Psi_6\rangle]$}}
 \begin{tabular}{lll}
  \hspace{-0.5cm}
  \includegraphics*[width=0.45\textwidth]{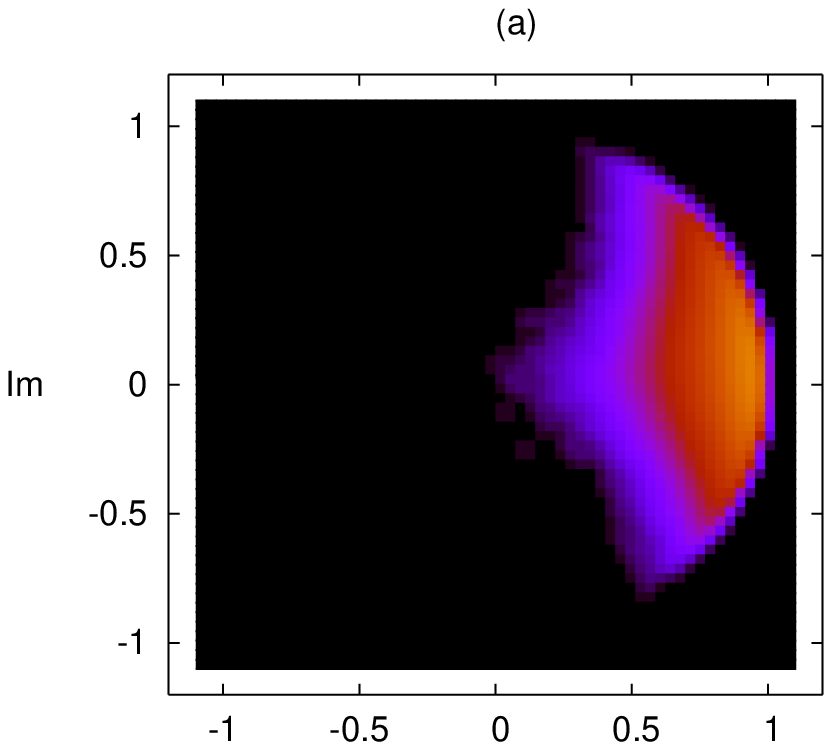} &
  \hspace{-2.4cm}
  \includegraphics*[width=0.45\textwidth]{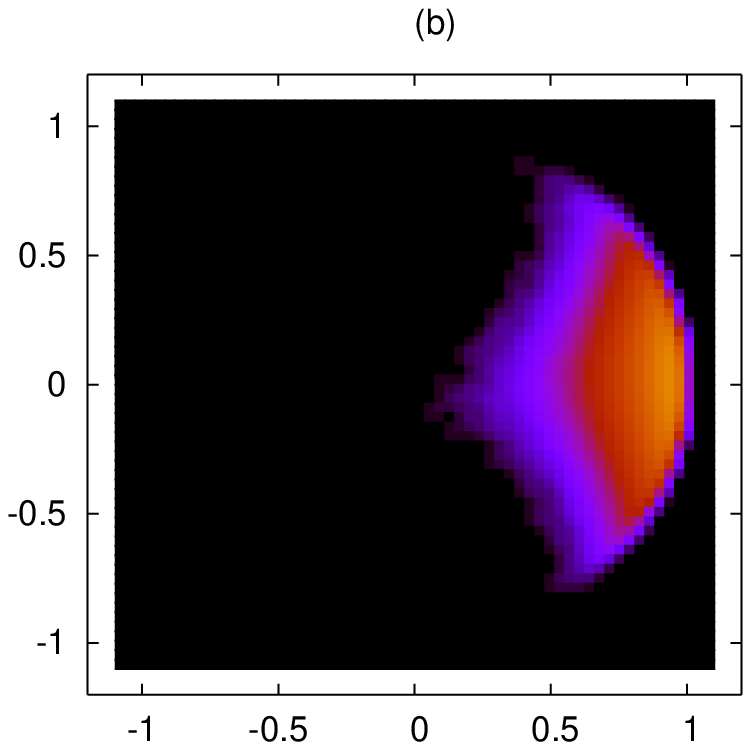} &
  \hspace{-2.4cm}
  \includegraphics*[width=0.45\textwidth]{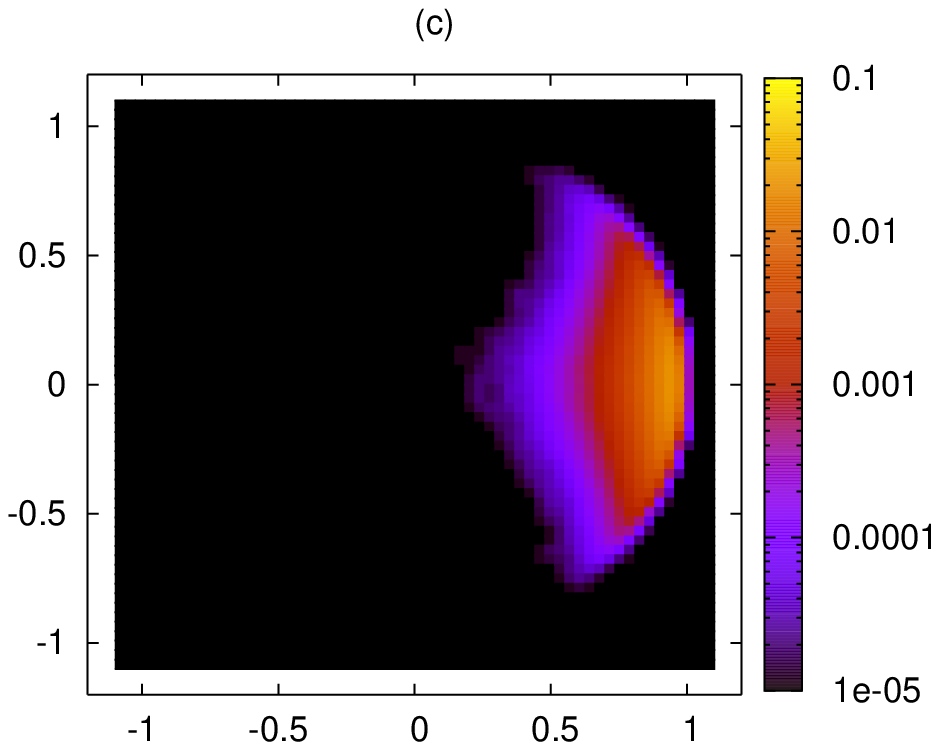} \\[-2.2cm]
  \hspace{-0.5cm}
  \includegraphics*[width=0.45\textwidth]{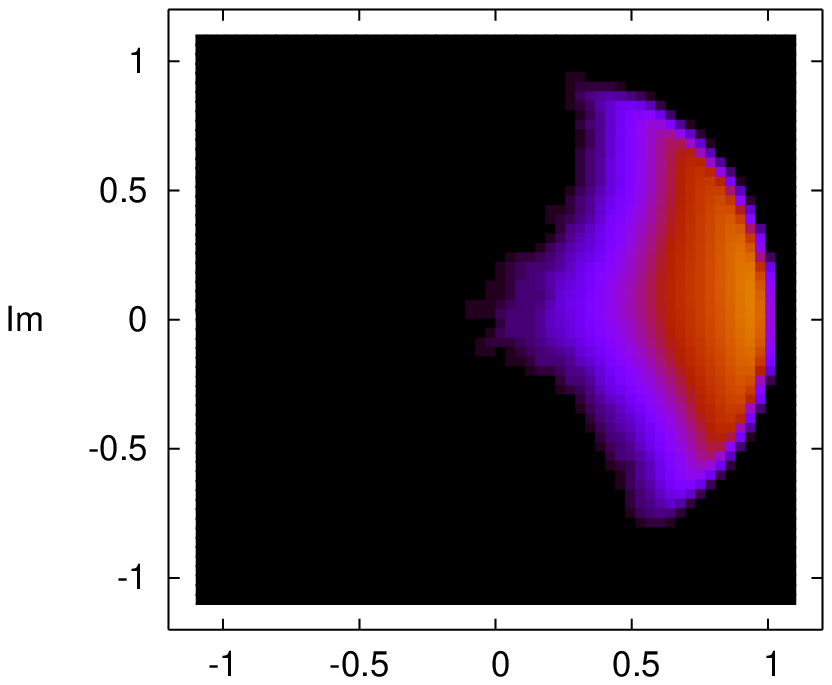} &
  \hspace{-2.4cm}
  \includegraphics*[width=0.45\textwidth]{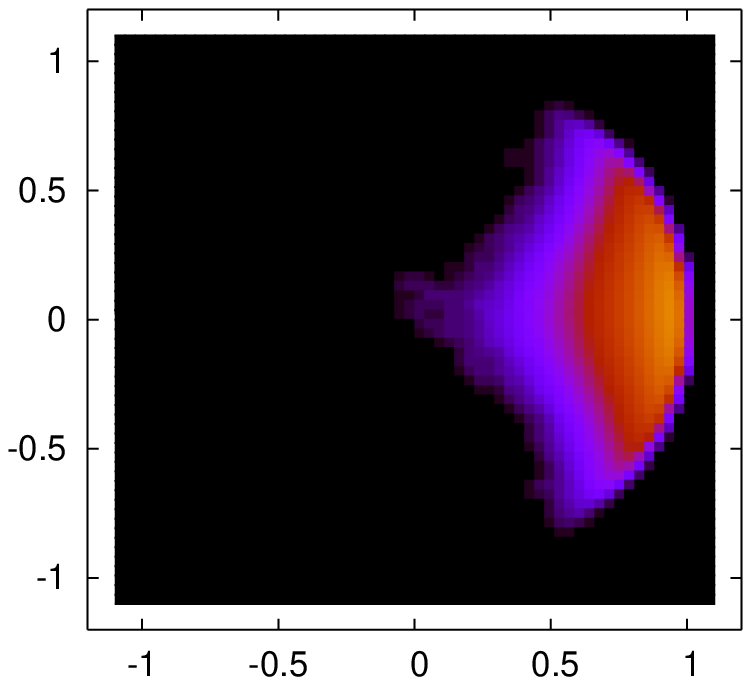} &
  \hspace{-2.4cm}
  \includegraphics*[width=0.45\textwidth]{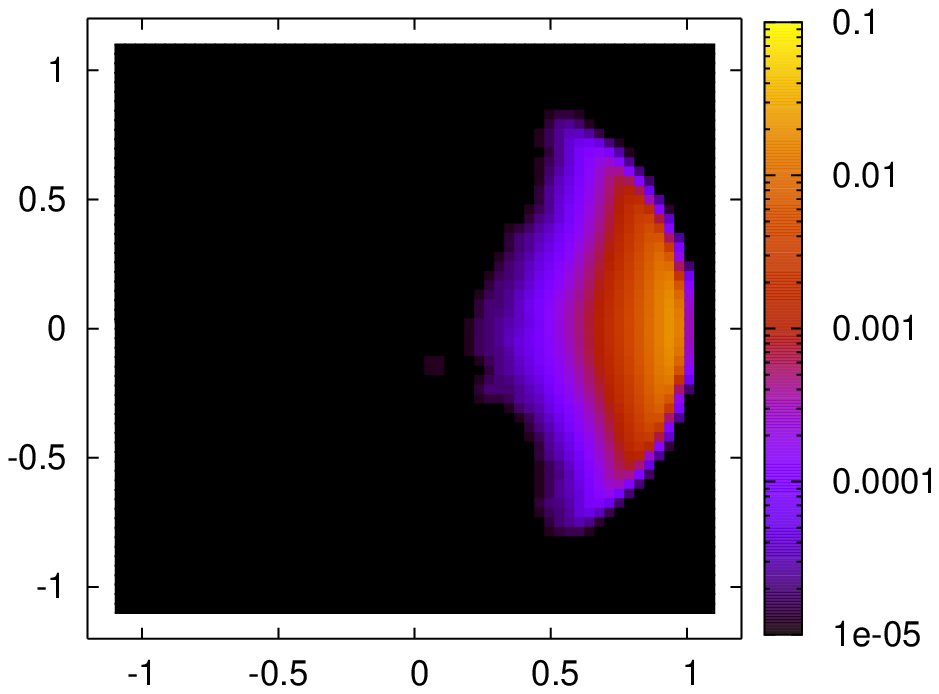} \\[-2.2cm]
  \hspace{-0.5cm}
  \includegraphics*[width=0.45\textwidth]{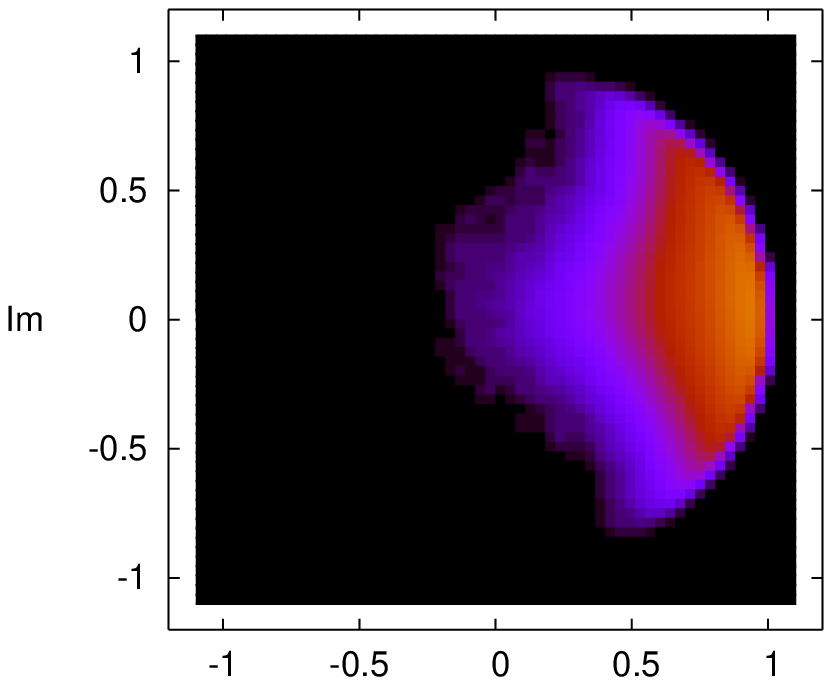} &
  \hspace{-2.4cm}
  \includegraphics*[width=0.45\textwidth]{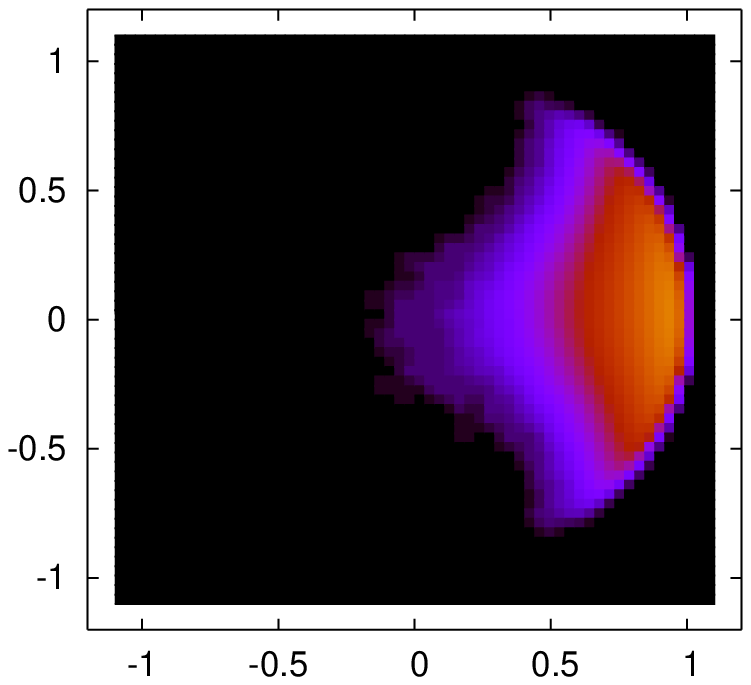} &
  \hspace{-2.4cm}
  \includegraphics*[width=0.45\textwidth]{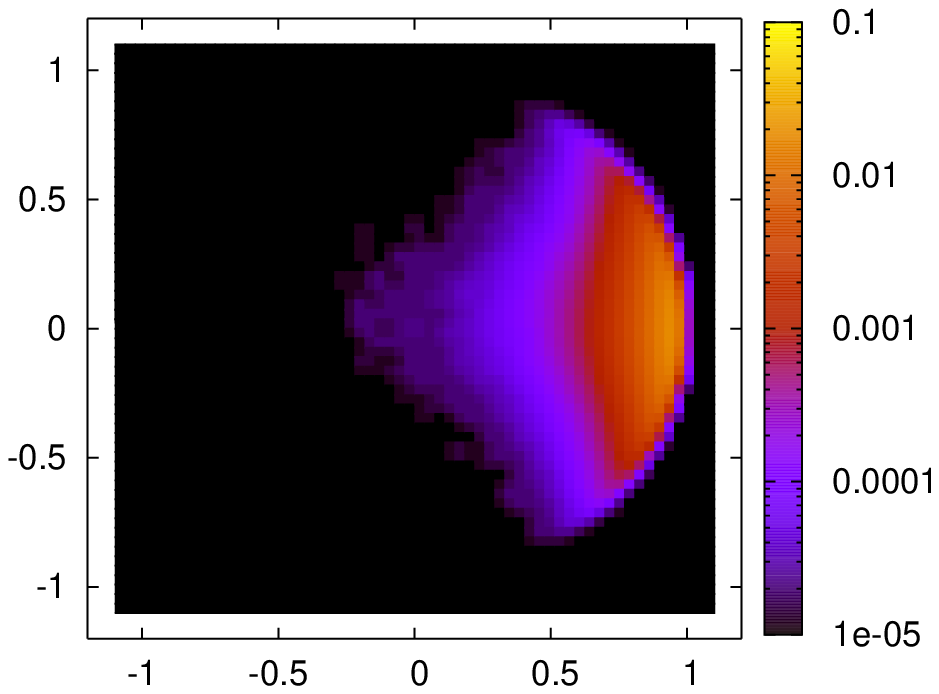} \\[-2.2cm]
  \hspace{-0.5cm}
  \includegraphics*[width=0.45\textwidth]{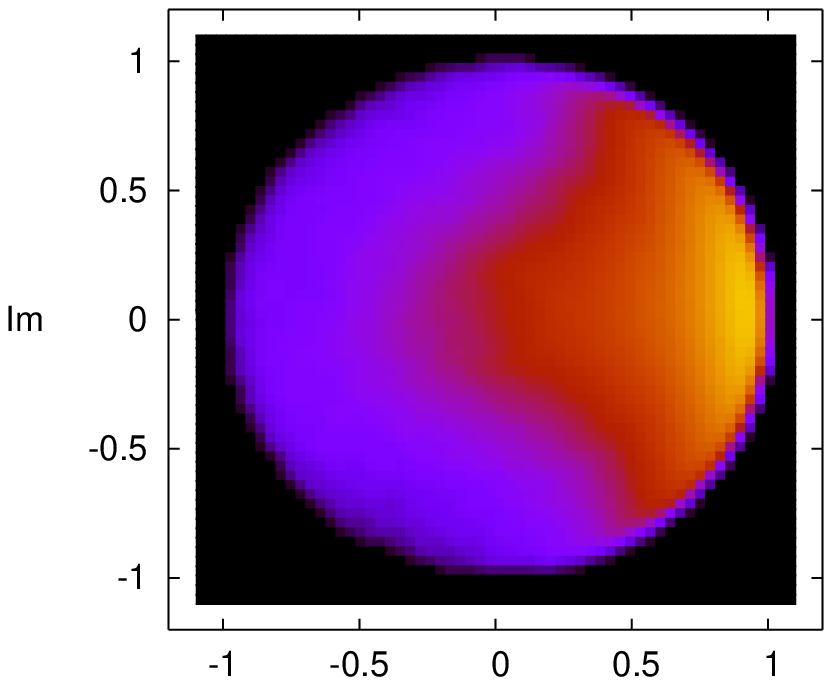} &
  \hspace{-2.4cm}
  \includegraphics*[width=0.45\textwidth]{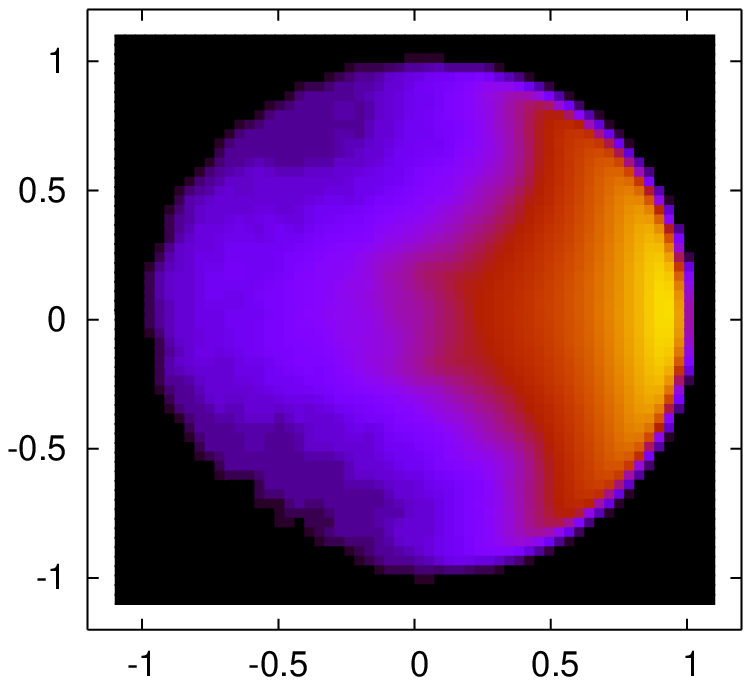} &
  \hspace{-2.4cm}
  \includegraphics*[width=0.45\textwidth]{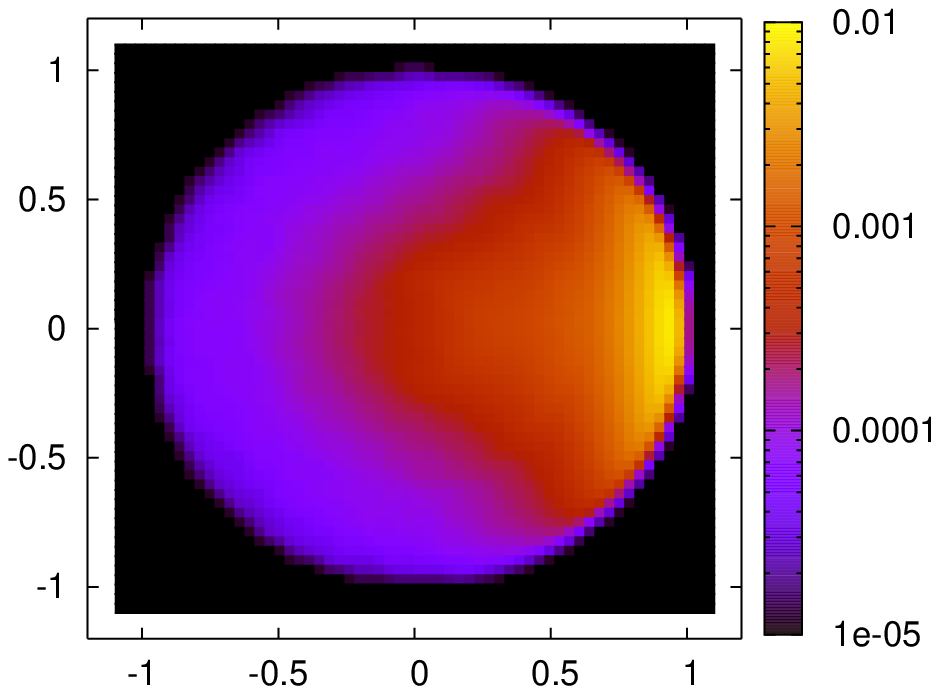} \\[-2.2cm]
  \hspace{-0.5cm}
  \includegraphics*[width=0.45\textwidth]{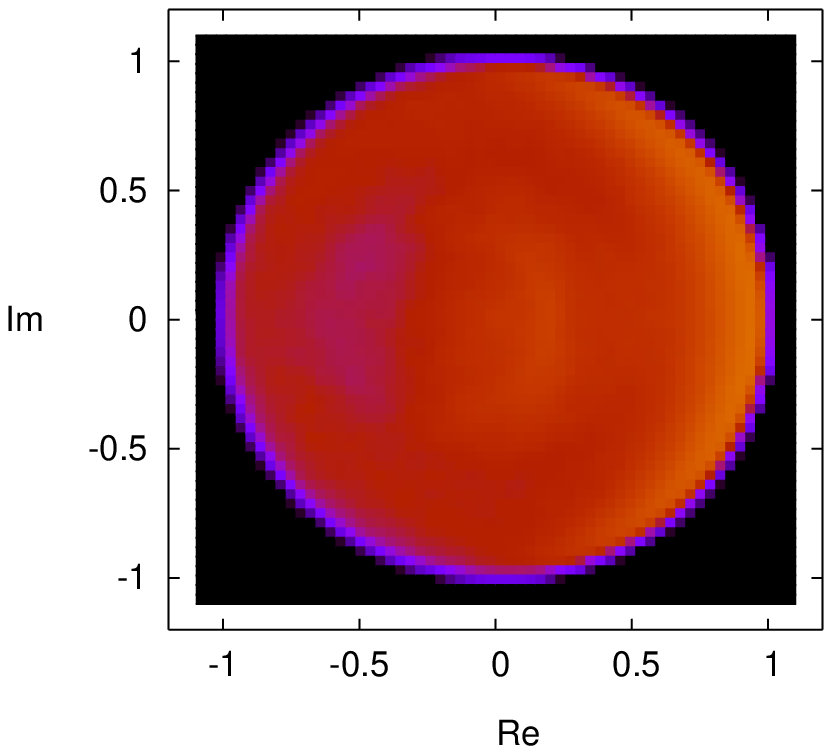} &
  \hspace{-2.4cm}
  \includegraphics*[width=0.45\textwidth]{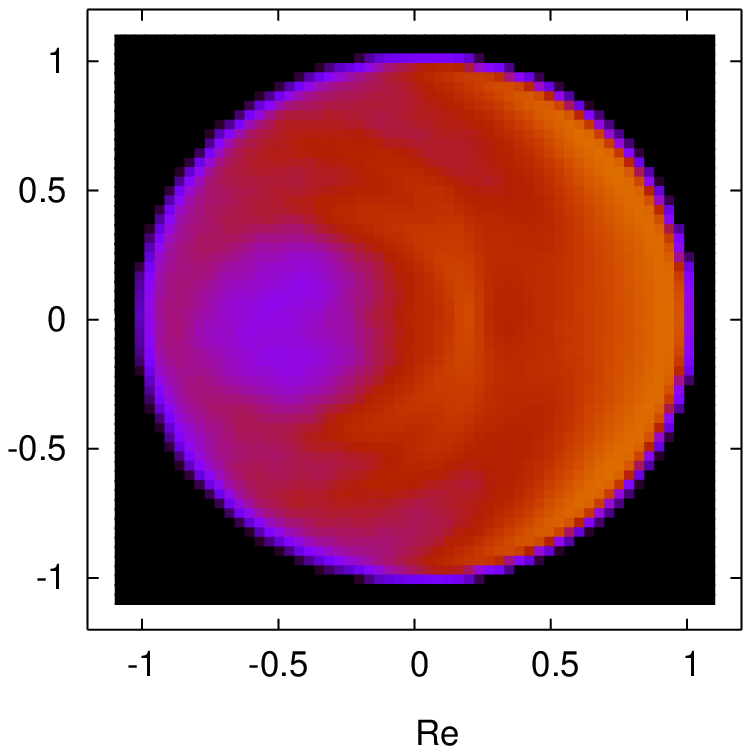} &
  \hspace{-2.4cm}
  \includegraphics*[width=0.45\textwidth]{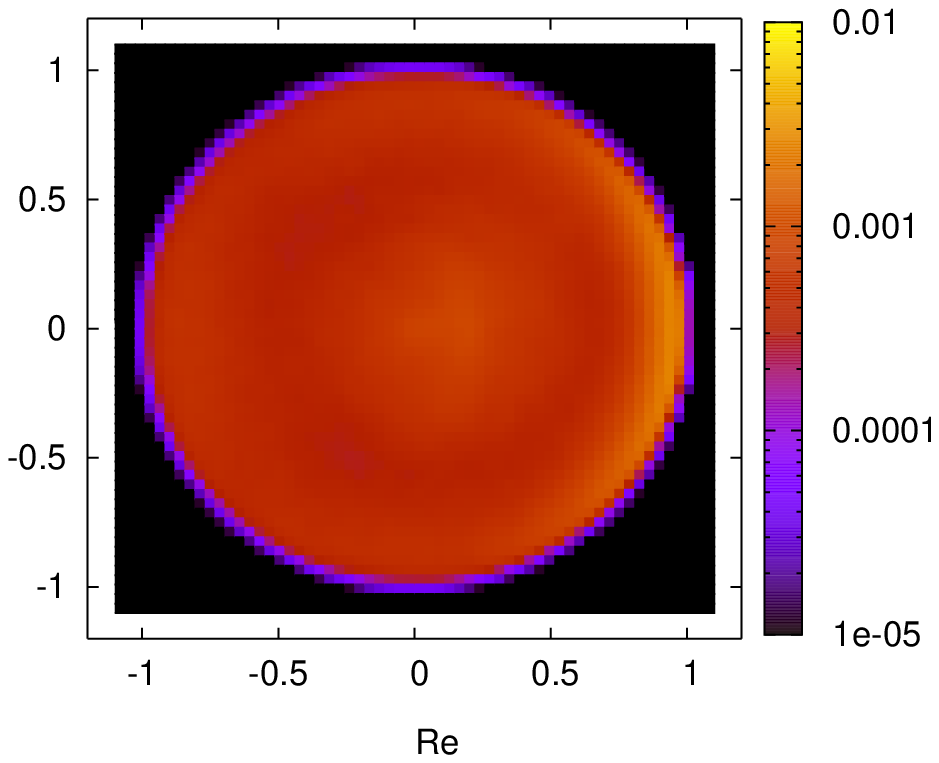} \\
 \end{tabular}
 \caption{(Color online) Intensity plots of the probability of $\Psi_6$ on the complex plane
          computed for different annular shells $(r_{min}\le r < r_{max})$
          for 2D confined $^4$He crystals with $R=54.6$ \AA. (column a) $N=671$ 
          (column b) $N=N_0(\rho_S)=685$ (column c) $N=699$;
          (first row) $0  \le r < 20$ \AA~
          (second row) $20 \le r < 30$ \AA~
          (third row) $30 \le r < 40$ \AA~
          (fourth row) $40 \le r < 50$ \AA~
          (fifth row) $50 \le r < R$ \AA. Note the logarithmic intensity scale.}
 \label{fig4}
\end{figure*}

Also the study of the orientational order parameter $\Psi_6$ corroborates the finding that no defects permeate
into the inner region of the confined crystal.
From the plots Fig.~\ref{fig4} it is evident that within 30 \AA~ from the center the results
for $\langle\Psi_6\rangle$ in the confined system with $R=54.6$ are close to the results obtained
in the periodically repeated ideal crystal reported in Fig.~\ref{fig3} for the three numbers of particles,
$N=671$, 685 and 699. 
In the range 30--40 \AA~ there is a crossover to the outer region of width $\sim 15$ \AA~ strongly perturbed
by the confining potential. 

In conclusion, our study of a confined system shows that the inner region is free of defects
like vacancies or interstitials; if defects are present in the initial configuration these 
anneal away and go to the interfacial region.
Since our system is finite, our computation allows to give only an upper bound to the
concentration of ground state defects.
Taking into account that the inner region has a radius of 35 \AA~ in which the average number
of $^4$He atoms is about 300, this upper bound is $3\times10^{-3}$.
We notice that this upper bound is above the estimated concentration of vacancies given by
a SWF (in 3D) \cite{Ross}.

\section{Conclusions}
\label{sec:conc}

The issue of ground state defects in crystalline $^4$He is still debated.
In 3D we have studied up to 98 vacancies in system of 2548 atoms and in 2D up to 30
vacancies in a system of 1410 atoms and in all cases the crystalline order, as studied
by different methods, turns out to be stable as long as the concentration of vacancies
is below about 2.5\%.
Multiple vacancies are {\it not} weakly interacting though, they display signature of strong
interaction.
The most detailed study has been performed in 2D.
For up to 6 vacancies one find a tendency for the vacancies to form linear structures.
Starting from 10 vacancies, in the relaxed state such vacancies cannot be identified anymore
but the system displays the presence of fluctuating dislocations.
This suggests that should ground state defects be present in solid $^4$He, such defects 
would be dislocations and not vacancies. 

We have stressed the role of periodic boundary conditions in quantum simulations of a
crystalline solid.
The computed quantities are affected not only by the usual size effects, but also by 
commensuration effects between the simulation box and the crystalline lattice.
We have investigated solid $^4$He in 2D without the use of PBC by confining the atoms
via an external potential.
We find that the effects of the confining potential extend mainly in a layer of width of order 
of 15 \AA, inside this interfacial region the atoms display a crystalline order similar
to that of a bulk crystal and we find no evidence for the presence of defects.
If defects like vacancies or interstitials are introduced in the initial state, such 
defects migrate to the interfacial region leaving the inner region free of defects.
Taking into account the finite size of the system, this allows us to put an upper limit of
$3\times10^{-3}$ to the concentration of zero--point defects.

Many extensions of the present study can be foreseen.
A stronger bound on the concentration of zero--point defects will require the study of
larger systems.
The walls of a real container are never smooth and it will be interesting to see how the
properties of the systems are affected by assuming a confining potential not as smooth as the
one used in the present study.
The study of confined system in 3D should be interesting but it is hindered by the very
large number of particles needed in order to get relevant results.

\begin{acknowledgements}
This work was supported by Regione Lombardia and CILEA Consortium through a LISA
Initiative (Laboratory for Interdisciplinary Advanced Simulation) 2010 grant
[link:http://lisa.cilea.it].
\end{acknowledgements}


\begin{thebibliography}{}
\bibitem{note} Particle configurations at fixed imaginary time have very strong fluctuations due to zero--point motion;
               in order to average over such fluctuations, as particle position we take the center of mass
               of the configurations of a suitable number of inner time--slices of the polymer. In the reported calculations
               this corresponds to an imaginary time interval of about 0.9 K$^{-1}$ out of $2\tau = 1.55$ K$^{-1}$.
\end{thebibliography}


\end{document}